\documentclass{article}

\begin {document}

\title {Einstein's Curvature for Nonlocal Gravitation of Gesamt Energy Carriers}

\author {I.E. Bulyzhenkov\\ {\small Institute of Spectroscopy RAS, Troitsk, Moscow reg., Russia and} \\
 {\small Department of Physics, University of Ottawa, Ottawa, Ontario, Canada}}



\maketitle

 \begin {abstract}
The intrinsic metric symmetries for energy-momentum in warped space-time universally reinforce strict spatial flatness in the GR metric formalism. The passive/active energy-charge for the 1686, 1913, and 1915 gravitational laws maintains the universal free fall in non-empty material space of  nonlocal elementary (radial) energies.  The known planetary perihelion precession, radar echo delay, and gravitational light bending can be explained by the singularity-free metric solution without departure from Euclidean spatial geometry. Non-Newtonian flatspace precessions are expected for non-point orbiting gyroscopes exclusively due to the GR inhomogeneous time in the Earth's radial energy-charge. The self-contained SR-GR relativity of {\it gesamt} particle-field carriers of inertial energy relies on the Principle of Equivalence for geometrization of the $r^{-4}$ continuous particle without references to Newton's mass-to-mass attraction. The post-Newtonian  logarithmic potential for distributed particle densities is also the exact solution to Maxwell's equations with the analytical $r^{-4}$ electric charge density instead of the delta-operator density.  

 \end {abstract}


\section {Introduction}

In 1913, Einstein and Grossmann published the {\it Entwurf} metric formalism for the passive material point in a gravitational field, and in 1915 Einstein's equation for energy sources (reiterated by Hilbert's variations) accomplished the basic tensor approach to warped space-time with active-passive gravitational matter \cite {1}. This metric theory of gravity, known as General Relativity (GR), can operate fluently  with curved spatial
displacement $dl_{_N} = {\sqrt {\gamma^{_N}_{ij}dx^idx^j}} $ of a point mass $m_{_N}$  by accepting  the
Schwarzschild or Droste empty-space solutions \cite {2}  without specific restrictions on the space metric tensor $\gamma^{_N}_{ij}\equiv   g^{_N}_{oi}g^{_N}_{oj}(g^{_N}_{oo})^{-1} -
g^{_N}_{ij}$.
All GR solutions are related to the  space-time interval of the considered carrier N,
$ds_{_N}^2 \equiv  g_{\mu\nu}^{_N}dx^\mu dx^\nu =
  d\tau_{_N}^2 - dl_{_N}^2$, where 
the time element $d\tau_{_N} \equiv [g^{_N}_{oo}
 (dx^o + g^{-1}_{_Noo}g^{_N}_{oi}dx^i)^2]^{1/2}$ depends on  
the local pseudo-Riemannian metric tensor $g^{_N}_{\mu\nu}$ 
 and, consequently,
 on local gravitational fields. Hereinafter, $i = 1,2,3 $, $\mu = 0,1,2,3$, and the speed of light c = 1, for short.

We intend to analyze the time, $d\tau_{_N}$, and space, $dl_{_N} \equiv {\sqrt {\gamma^{_N}_{ij}dx^idx^j}}$, elements within the conventional GR four-interval $ds  \equiv {\sqrt {g^{_N}_{\mu\nu}dx^\mu dx^\nu}}$ and to prove that the time element of the freely moving mass $m_{_N}$ depends within gravitational fields not only on Newton's absolute time $t$ (with $dt \equiv {\sqrt {\delta_{oo}dx^odx^o}}$  = $|dx^o| > 0$), but also on the absolute space coordinates $x^i$. Then the ratio $dl_{_N}/d\tau_{_N} \equiv v$ (the physical speed in  Special Relativity) will non-linearly depend on spatial displacement $dl_{_N} \equiv {\sqrt {\gamma^{_N}_{ij}dx^idx^j}}$ (the space interval in Special Relativity). Non-linear field contributions to the time element $d\tau_{_N}(v)$ within the conventional four-interval  $ ds^2$ = $
  d\tau^2 (v) - dl^2$ of Einstein's Relativity may modify Schwarzschild-type metric constructions with curved three-space around non-physical point singularities of GR energy-sources. Moreover, the calculated ratio $dl_{_N}/d\tau_{_N} (v) = v$  may differ from a real velocity $dl_{_N}/d\tau_{_O}$ measured by a motionless local observer with proper-time $d\tau_{_O} (dl=0) \neq d\tau_{_N} (v)$. This is most evident for the gravitational Sagnac effect when $g_{oi}dx^i/d\tau \neq 0$. We expect that carrier's non-linear time $d\tau (v) \equiv d\tau (dl)$ may preserve the universal flatspace element $dl$ even in a strongly warped space-time interval $ds$.   We shall start within  the framework of the 1913 tensor formalism for the geodesic motion of passive mass-energies. However, our non-linear solution for the GR interval can also be found from full geometrization of active source-energy densities, if one generalizes the Einstein tensor curvature on continuous particles and their fields when ${\hat G}_o^\mu = 0$ and ${\hat G}_{o;\mu}^\mu \equiv 0$ (or if one drops {\it ad hoc} empty space and point source dogmas in favor of Newton's and Clifford's material space and the 1913 {\it Entwurf} metric theory with 3D flatspace).

The first post-{\it Entwurf} attempt to interpret GR in parallel terms of curved and flat four-spaces was made by Rosen 
\cite {3}, Einstein's co-author of the unpublished 1936 paper about the non-existence of cylindrical metric waves from the `coordinate' source singularities in question. Later, Sommerfeld, Schwinger, Brillouin and many other theorists tried to protect Euclidean space for modern physics, including quantum electrodynamics and light waves. Originally,  Einstein-Grossmann's idea for the geometrization of gravitational fields also relied in the 1913 {\it Entwurf}  project on flatspace gravitation. Contrary to non-metric approaches to gravitation with spatial flatness,  for example \cite {4}, we shall comply with the Einstein-Grossmann extension of Special Relativity (SR) to gravitation through  warped space-time with  non-Euclidean pseudo-geometry, developed by Lobachevsky, Bolyai, and Riemann \cite {Lbg}. However, the coherent 4D geometrization of  fields and particles may keep intrinsic metric symmetries  $\gamma^{_N}_{ij} = \delta_{ij} $ (or universal 3D subgeometry) in the GR tensor formalism for every physical object. In other words, we are planning to revise neither Einstein's Principle of Relativity nor the GR metric concept. On the contrary, we are planning further GR geometrization of continuous particles-sources beside the already developed geometrization of gravitational fields. Full nullification of the continuous Einstein curvature for overlapping local densities of distributed (astro)particles and their fields will be required. We intend to prove, for example, that Schwarzschild's solution for a central field is not `the only rotationally invariant GR metric extension of the SR interval' if one admits non-empty material space or Newtonian stresses of the material medium-ether 
associated with continuous distributions of non-local gravitating bodies. 

First, we discuss a local time element, $d\tau \equiv d\tau (dl, v)$, which should be considered in GR as a non-linear function of the speed $v = dl/d\tau$ or spatial displacement $dl$ of a passive material point in a given gravitational field. Then, we look at  the Weber-type potential $U_o^{_W}=U_o{\sqrt {1-v^2}}/m_{_N} = U_oP^{-1}_o /(1 - U_oP_o^{-1})$ for a  planet with mass $m_{_N}$ and  relativistic energy $P_o = m_{_N}V_o$ in the Sun's static field generated by the active energy-charge $E_{_M} = MV_{_Mo}$. Ultimately, we derive the energy-to-energy attraction potential $U_o/P_o = - GE_{_M} /r \equiv - r_o/r$ for global Machian interactions of nonlocal GR particles with an analytical radial density $n(r)  = r_o/4\pi r^2 (r + r_o)^2 $ instead of the delta operator density from the conventional approximation of matter through localized point particles in field spaces. We then find arguments for the singularity-free gravitational contribution $U_o^{_W}= - r_o/ (r + r_o)$ to the smooth metric tensor component ${{g_{oo}}} = (1 + U_o^{_W})^2$. The main challenge here is to keep the free fall universality and the GR Principle of Equivalence for all carriers of probe (passive, inertial) energies $P_o$ in radial fields of the Sun's active (attractive) energy $E_{_M}$.  

In our approach, the warped GR four-interval $ds[d\tau(dl,v), dl]$ cannot be decomposed conceptually into pure time and space subintervals, contrary to the algebraic Schwarzschild-type solutions \cite {2} with the time and space metric split. In order to justify the indivisible non-linear composition of time and space elements in the GR four-interval we clarify how the already known gravitational tests of GR can be explained quantitatively without departure from spatial flatness.
We finalize astrophysical tests of our energy-to-energy attraction under the Einstein-Grossmann geodesic motion in metric fields without Schwarzschild singularities by pro-Einstein-Infeld-Hoffmann comments on non-point slow-moving gyroscopes for the Gravity Probe B.

We  verify the {\it Entwurf} hypothesis that the GR metric generalization should universally admit 
flatspace metric symmetries $\gamma^{_N}_{ij} = \delta_{ij} $ for every elementary particle N with energy  $P_o = m_{_N} {\sqrt {g_{oo}}}/ {\sqrt {1-v^2}}$ in warped space-time. Then, we find that the metric tensor $g_{\mu \nu }$ with four gravitational potentials $G_\mu \equiv U_\mu/P_o$ for the particle energy $P_o$ can satisfy these metric symmetries for non-empty material space under any gravitational fields and their gauges. This neo-{\it Entwurf} metric scheme with warped space-time, but strictly flat three-space, is consistent with the Universe's large-scale flatness confirmed by the balloon measurements 
\cite {5} of the nearly isotropic 2.73K cosmic microwave background radiation. This scheme, non-linearly dilated time in material flatspace, explains quantitatively the planet perihelion precession, radar echo delay, and gravitational light bending, for example \cite {6}.

Our metric corrections to Newtonian motion in the Sun's weak field coincide with  similar results of other authors who traditionally admit empty curved space and decompositions of the invariant four-interval into its algebraic time and space parts. Observable dynamics of matter in moderate and strong fields provides, in principle, an opportunity to distinguish GR solutions with non-linear time and flat space from Schwarzschild-type constructions without intrinsic metric symmetries. Precise non-relativistic experiments in the Earth's gravitational field can verify the nonlocal source geometrization with non-linearly dilated time, flat material space, and the $r^{-4}$ particle-field carrier of energy in the self-contained gravitation with global energy-to-energy attraction and local energy-driven 4D geometry. 

\section {Warped four-space with intrinsic metric symmetries for flat three-space} 

To begin,  we employ the GR tetrad formalism,
 for example \cite {8,9}, in covariant expressions for an elementary rest-mass $m_{_N}$ in order to justify the universal mathematical opportunity to keep a
flat 3D subspace $x_{_N}^i$ in curved four-space $x^\mu_{_N}$
 with a pseudo-Riemannian metric tensor $g^{_N}_{\mu \nu} = g_{\mu \nu }$ (for short).
First, we rewrite the four-interval,
  $ds_{_N}^2 \equiv g^{_N}_{\mu \nu }dx_{_N}^\mu dx_{_N}^\nu $ $\equiv$ 
$g_{\mu \nu }dx^\mu dx^\nu $
 $\equiv \eta_{\alpha \beta  }e^{(\alpha)}_{\ \mu} e^{(\beta)}_{\ \nu} dx^\mu
  dx^\nu $ $\equiv \eta_{\alpha \beta}dx^{(\alpha)} dx^{(\beta)} $,
in plane coordinates
 $dx^{(\alpha)}
\equiv e_{\ \mu} ^{(\alpha)} dx^\mu $ and $dx^{(\beta)}
\equiv e_{\ \nu} ^{(\beta)} dx^\nu $, with
 $\eta_{\alpha \beta }$ 
= $diag (+1,-1,-1,-1)$. One can  find
$e^{(o)}_{\ \mu} = \{  {\sqrt {g_{oo}}} ; - {\sqrt {g_{oo}}}g_i  \}$
and $e^{(b)} _{\ \mu} = \{0, e_{\ i}^{(b)}   \} $ from the equality
$ds^2 \equiv [{\sqrt {g_{oo}}} (dx^o-g_idx^i)]^2 - \gamma _{ij}dx^idx^j$,
$g_i \equiv - g_{oi}/g_{oo}$.
At first glance, the spatial triad $e_{\ i}^{(b)}$ $\equiv$ $e_{_N i}^{(b)}$
 (a, b = 1,2,3 and $\alpha, \beta$ = 0,1,2,3) depends essentially on the gravitational
  fields of other particles because this triad is related to components of $g^{_N}_{\mu\nu}$.
  However, this might not be the case when there are certain intrinsic 
symmetries for the pseudo-Riemannian metric with the warped tensor $g^{_N}_{\mu\nu}$.
Shortly, a curved mathematical 4D manifold does not necessarily mean a curved 3D subspace for real matter (warped papers in 3D trash, for example,  keep parallel lines of Euclidean 2D geometry).

Let us consider three spatial components $V_i$ of the  four-vector
$V_\mu\equiv g_{\mu\nu}{dx}^\nu/ds$
 by
using the conventional tetrad formalism, $-({\sqrt {g_{oo}  }}g_i + v_i  )(1 -
v_iv^i)^{-1/2} $ $\equiv$
  $V_i \equiv e^{(\beta)} _{\ i} V_{(\beta)} \equiv e^{(o)}_{\ i} V_{(o)} + e_{\ i}^{(b)} V_{(b)} $
$\equiv$ $-({\sqrt {g_{oo}  }}g_i + e_{\ i}^{(b)}v_{(b)}  )(1 - v_{(b)}v^{(b)})^{-1/2} $.
Here, we used $e^{(o)}_{\ i} =  - {\sqrt {g_{oo}}}g_i $ and $V_{(\beta)} = \{
(1-v_{(b)}v^{(b)})^{-1/2}; - v_{(b)}(1-v_{(b)}v^{(b)})^{-1/2} \}$. Now one can  trace that the considered equalities $V_i \equiv e^{(\beta)} _{\ i} V_{(\beta)}$ can admit trivial relations $v_iv^i = v_{(b)}v^{(b)}$ and $v_i = e_{\ i}^{(b)}v_{(b)} = \delta_i^{(b)}v_{(b)}$ between the curved velocities,  $v_i \equiv \gamma_{ij}dx^j /{\sqrt
 {g_{oo}}}(dx^o - g_i dx^i) \equiv \gamma_{ij}dx^j/d\tau$,
and the plane velocities, $v_{(b)} = \delta_{ab}dx^{(a)}/d\tau $.  
These `trivial' relations indicate that all spatial triads
can be considered as universal  Kronecker delta symbols, $e_{_N i}^{(b)} = \delta _{\ i}^{(b)}$, and, consequently, the three-space metric
tensor  is irrelevant to gravitation fields, {\it i.e.}  
$  g_{oi}g_{oj}g^{-1}_{oo} - g_{ij}$ $\equiv$ $\gamma_{ij} = \gamma^{_N}_{ij} = \gamma^{_K}_{ij}$ = $\delta _{ij}$. 

Again, we can read $ g^{_K}_{\mu \nu } \equiv
\eta_{\alpha \beta} e^{(\alpha)}_{\ \mu} e^{(\beta)}_{\ \nu}$, with
$e^{(o)}_{\ \mu} = \{  {\sqrt {g_{oo}}} ; - {\sqrt {g_{oo}}}g_i  \}$
and $e^{(b)}_{\ \mu} = \{0, \delta _{\ i}^{(b)}   \} $ $\equiv \delta_{\ \mu}^{(b)} $
in the most general case. From here $g_{oo} \equiv e^{(o)}_{\ o} e^{(o)}_{\ o}$, $g_{oi} \equiv e^{(o)}_{\ o} e^{(o)}_{\ i}$, 
and $g_{ij} \equiv e^{(o)}_{\ i} e^{(o)}_{\ j} - \delta_{ab}e^{(a)}_{\ i}e^{(b)}_{\ j} = e^{(o)}_{\ i} e^{(o)}_{\ j} - \delta_{ij}$. 
Therefore, Euclidean spatial geometry can be universally applied by the 
 covariant GR formalism to  $dl_{_K}^2 \equiv \gamma^{_K}_{ij} dx^i dx^j = \delta_{ij} dx^i dx^j$ in pseudo-Riemannian metrics due to the intrinsic symmetries $\gamma^{_K}_{ij} 
 \equiv    g^{_K}_{oi}g^{_K}_{oj}(g^{_K}_{oo})^{-1} - g^{_K}_{ij}$ 
   $\equiv$  $ \delta_{ij}$.  
 
 Contrary to universal spatial displacements $dl$, invariant four-intervals have differently warped pseudo-Riemannian geometries for particles K and N,  because $g^{_N}_{\mu\nu} \neq g^{_K}_{\mu\nu}$ and $ds_{_K} \neq ds_{_N}$    
 in different external fields (for example, in the two-body problem).
The GR four-interval for a selected mass-energy carrier, 
\begin {equation}
ds^2\equiv d \tau^2  - dl^2 
 = \left ( {\sqrt {g_{oo}}}dx^o  + {{g_{oi}dx^i}\over{\sqrt {g_{oo}}} }  \right )^2
 - {\gamma_{i j} dx^i dx^j}, 
\end {equation}
 is defined for only one selected probe mass $m_{_N}$; $ ds_{_N} \equiv ds$ and $dx_{_N}\equiv dx$ are used exclusively for brevity.  This interval cannot be rigorously divided into time, $d \tau^2 $, and space, $dl^2 \equiv  {\gamma_{i j} dx^i dx^j} = {\delta_{i j} dx^i dx^j}$, elements. We  prove below that $d \tau \equiv d\tau(dl)$ for constant gravitational fields with the first integral of motion $P_o = const$. Such a time element $d\tau \equiv d\tau_{_N}(dl) \equiv {\sqrt {g_{oo}}} (dx^o - g_i dx^i)$  of the moving mass $m_{_N}$
 always counts its spatial displacement $dl$ in a gravitational field,  despite the fact that  it not immediately obvious from the GR time definition for fields with $g_{oi} = 0$. This post-Newtonian phenomenon appears for energy(velocity)-dependant potentials in  nonlinear gravitational equations. Our interpretation of the warped four-interval (1), based  on  energy-warped time in non-empty flatspace rather than in empty warped space, may be considered as an alternative way in Einstein's relativity for  unified geometrization of a {\it gesamt} elementary carrier, which is a distributed radial field together with a distributed radial particle.

Now, we return to components of the four-vector $V^{_N}_{\mu}=g^{_N}_{\mu\nu}dx^\nu/ds$.  Notice that
$V_\mu = e^{(\alpha)}_{\ \mu} V_{(\alpha)}  = (e^{(b)}_{\ \mu} V_{(b)} + e^{(o)}_{\ \mu} V_{(o)})$
= $ (e_{\ \mu}^{(b)} V_{(b)}  + \delta^{(o)}_{\ \mu} V_{(o)} ) $
+ $ (e_{\ \mu}^{(o)} - \delta^{(o)}_{\ \mu} ) V_{(o)}  $
$\equiv$ ${\cal V}_\mu + m^{-1}_{_N}U_\mu$,  with the four-velocity   
$ {\cal V}_\mu  $ $\equiv$
$e_{\ \mu}^{(b)} V_{(b)}  + \delta^{(o)}_\mu V_{(o)} = \delta_\mu^\alpha V_\alpha$,
 because $e^{(b)}_{\ o} = 0$ and $e^{(b)}_{\ i} = \delta^{(b)}_i$.
Flat three-space geometry
is a promising way  to introduce gauge invariant gravitational potentials,  $G_\mu \equiv U_\mu / P_o = G'_\mu + \partial_\mu \phi_{_N}$ with 
 $U_\mu  \equiv (e_{\ \mu} ^o   - \delta^o_\mu  )m_{_N}V_o $ $= U'_{\mu}  
 + P_o\partial_\mu \phi_{_N}$, for the Einstein-Grossmann `material point', in close analogy to four-component electromagnetic potentials for the classical electric charge.
 The point is that a four-momentum $P^{_N}_{\mu } \equiv m_{_N}V^{_N}_{\mu}$  of the selected scalar mass $m_{_N}$ can be
  rigorously decomposed into mechanical, $K^{_N}_\mu$, and  gravitational, $U^{_N}_\mu$, parts
  only under strict spatial flatness, 
  \begin{eqnarray} 
  P^{_N}_{\mu}  
=    \!\left \{\!{{m_{_N} }\over {\sqrt {1 - v^2}}};\!-{{m_{_N} v_i}\over {\sqrt {1 - v^2}}}\right \} 
   + \!\left \{\!{{m_{_N} ({\sqrt {g_{oo}}}-1)  }\over {\sqrt {1-v^2}}};-
  {{m_{_N}g_i{\sqrt{g_{oo}}} }\over {\sqrt {1 - v^2}}}\right \} 
 \equiv K^{_N}_\mu + U^{_N}_\mu ,
  \end{eqnarray}
where $v_i\equiv \gamma_{ij}v^j, v^2 \equiv v_iv^i$,
$v^i\equiv $  $dx^i /d\tau$,
$ds$  =  $(dx_\mu dx^\mu)^{1/2}$, $dx_{\mu} = g_{\mu\nu} dx^\nu$,
$dx^\mu \equiv dx^\mu_{_N}$, $ g_i$ $\equiv$ $-g_{oi}/g_{oo}$;  $\gamma_{ij}\equiv g_ig_jg_{oo}-g_{ij} = \delta_{ij} = - \eta_{ij}$. Again, we use for simplicity only one time-like worldline with $dt = dx^o > 0$ and $d\tau = +g_{oo}^{1/2}(dx^o-g_idx^i)> 0$ for the passive-inertial $m_{_N}> 0$, while real chiral matter should employ both time coordinates in the joint time arrow $dt = |\pm dx^o|$.
The gravitational energy-momentum part $U_\mu$ is defined in (2) for a selected mass $m_{_N}$ and its positively defined passive energy $P_o = m_{_N} V_o > 0$, associated  with the global distribution of all other masses $m_{_K}$. This gravitational part, $U_\mu \equiv G_\mu P_o$,  is not a full four-vector in pseudo-Riemannian space-time, like 
 $P^{_N}_\mu$, nor is the mechanical summand $K_\mu \equiv m_{_N}{\cal V}_\mu$.

 Because  $e_{\ \mu} ^{(b)} = \{0,   \delta_{\ i}^{(b)} \}$
$= \delta^{(b)}_{\ \mu}$ and $dx_\mu =  e_{\ \mu} ^{(\beta)}dx_{(\beta)}$,  the tetrad with the 
zero ({\it i.e.} time) label takes  the following
components from (2): $e^{(o)}_{\ \mu} =
\{1 +   {\sqrt {1-v^2} } U_om^{-1}_{_N};
 {\sqrt {1-v^2} }U_i m^{-1}_{_N} \} $
 $= \delta^{(o)}_\mu +  {\sqrt {1-v^2} }U_\mu m^{-1}_{_N} $. 
  Finally, the tetrad $e_{\ \mu} ^{(\beta)}$ for the selected particle $N$ and the metric tensor $g^{_N}_{\mu \nu } \equiv \eta_{\alpha\beta}e^{(\alpha)}_{\ \mu} e^{(\beta)}_{\ \nu}$, with $g_{\mu\nu}g^{\mu\lambda} = \delta^\lambda_\nu$,    
  depend only on the gravitational four-potential $U_\mu/P_o \equiv G_\mu$ for passive energy-charges $P_o \equiv P_{o{_N}}$,
\begin{equation}
\begin {cases}  
{e_{\ \mu}^{(\beta)} = \delta^{(\beta)}_{\ \mu} + \delta^{(\beta)}_{\ o} {\sqrt {1-v^2} }U_\mu m^{-1}_{_N} = \delta^{(\beta)}_{\ \mu} + \delta^{(\beta)}_{\ o} U_\mu P^{-1}_o/ (1 - U_oP^{-1}_o) \cr
\cr
g^{_N}_{oo} \equiv e^{(o)}_{\ o}e^{(o)}_{\ o} = (1 +   {\sqrt {1-v^2}} U_o m^{-1}_{_N})^2 
= 1/(1 -U_oP^{-1}_o)^2
  \cr \cr
       g^{_N}_{oi} \equiv e^{(o)}_{\ o}e^{(o)}_{\ i} = (1 +   {\sqrt {1-v^2} }U_o m^{-1}_{_N}) {\sqrt {1-v^2} }U_im^{-1}_{_N} = 
        { {g^{_N}_{oo}}}U_iP^{-1}_o \cr
 \cr g^{_N}_{ij} \equiv e^{(o)}_{\ i}e^{(o)}_{\ j} -  \delta_{ab}e^{(a)}_{\ i}e^{(b)}_{\ j} =({1-v^2})U_i U_jm^{-2}_{_N}   - \delta_{ij} = g^{_N}_{oo}U_iU_jP^{-2}_o  - \delta_{ij}\cr\cr
 g_{_N}^{oo} = (1-U_oP^{-1}_o)^2 - \delta^{ij}U_iU_jP_o^{-2}, \ g_{_N}^{oi} = U_iP^{-1}_o, \ g_{_N}^{ij} = - \delta^{ij},
  }
     \end {cases}
\end{equation}
where we used $g_{oo} \equiv e^{(o)}_{\ o}e^{(o)}_{\ o} = (1 +   {\sqrt {1-v^2}} U_o )^2$ and $V_o^2 = g_{oo}/ (1-v^2)$ to prove that ${\sqrt {g_{oo}}} =  1 +  {\sqrt {g_{oo}}} U_oP^{-1}_o  = 1/ (1 - U_oP^{-1}_o)$. Therefore, the passive-inertial GR energy, $P_o = m{\sqrt {g_{oo}}}/{\sqrt {1-v^2}} = m/{\sqrt {1-v^2}} (1 - U_oP^{-1}_o) = (m/{\sqrt {1-v^2}}) + U_o$, takes a linear superposition of kinetic and potential energies in  strong external fields. 
Note that we did not assign spin $S_\mu$ or other non-metric energy (heat) to the Einstein-Grossmann `material point' $m_{_N}$ with the translation energy-momentum (2). The affine connections for the metric tensor (3) depend  only on four gravitational potentials $U_\mu/P_o$ in our space-time geometry, which is not relevant to warped manifolds with asymmetrical connections and torsion fields \cite{Car}.

Every component of the metric tensor in (3) depends on the gravitational part $U_{\mu} \equiv m_{_N}V_\mu - m_{_N}{\cal V}_\mu \equiv G_\mu P_o$ of the probe carrier energy-momentum $P_\mu$. At the same time, all the components of the three-space metric
tensor,  $\gamma_{ij} \equiv g_{oi}g_{oj}g_{oo}^{-1} - g_{ij} = \delta _{ij}$, are always 
independent from the gravitational potential $G_\mu = U_{\mu}/P_o$ or its gauge. Such inherent metric symmetries for 3D subspace may be verified  directly from (3).
In fact, the GR tetrad and the metric tensor depends on the inharmonic Weber-type potentials, $ U^{_W}_\mu \equiv U_\mu{\sqrt {1-v^2}}/m_{_N} = U_\mu P^{-1}_o/ (1 - U_oP^{-1}_o)$, associated with the particle speed $v^2 = dl^2/d\tau^2$. In 1848 Weber  introduced \cite{Web}  the non-Coulombic potential $q_{_1}q_{_2} (1 - v^2_{_1{_2}}/2)/  r_{_1{_2}}$ based on lab measurements of accelerating forces between moving charges $q_{_1}$ and $q_{_2}$ with the relative radial velocity $v^2_{_1{_2}} \ll 1$. This was the first experimental finding that mechanical inertia and acceleration depend on the kinetic  energy of interacting bodies.       

By substituting the metric tensor (3) into  the interval $ds^2$ $\equiv$ $g_{\mu\nu}dx^\mu dx^\nu$ = $d^2\tau - dl^2$, one can rewrite (1) and find the proper time $d\tau =d\tau_{_N}$ of the probe mass-energy carrier N in external gravitational fields for the most general case,
\begin {equation}
 d\tau\equiv[g^{_N}_{oo}
 (dx^o+g^{-1}_{_Noo}g^{_N}_{oi}dx^i)^2]^{1/2}\equiv e^{(o)}_{\ \mu}dx^\mu=
 dx^o+{dx^\mu}{{U^{_N}_\mu}\over m_{_N}}{\sqrt{{{1-{dl^2\over d\tau^2(dl,v)}}}}}.
 \end {equation}

Notice  that the proper-time differential, $d\tau_{_O}=dx^o(1+U^{_K}_om^{-1}_{_K})$, of the  local observer $K$, with $ dx_{_K}^i = 0$ and $dl_{_K} = 0$, differs from the time element (4) of the moving mass $m$ with the GR energy-charge $P_o = m {\sqrt {g_{oo}}/ {\sqrt {1-v^2}}}$. 
The proper interval $ds$ of the moving mass and its proper time element (4) depends, in general, on all four components of $U_\mu$.  
Therefore, the observed three-speed $dl/d\tau_{_O}$, of a particle may differ
in relativistic gravito-mechanics from the non-linear ratio $dl/d\tau(dl,v) \equiv v$ of the particle's space and time elements of the invariant (1). 

 The metric tensor (3), the interval (1), and the local time element (4) are associated with warped space-time specified by external fields for one selected mass $m_{_N}$ or its passive energy-charge $P^{_N}_o$. One may employ common three-space for all elementary particles (due to universal Euclidean geometry for their spatial displacements), but  should specify warped space-times with differently dilated times for the mutual motion of gravitational partners. The particle's time element $d\tau \equiv d\tau_{_N} (dl,v)$ in (4) depends on the ratio  $U^{_N}_\mu/m_{_N}$ for one selected mass and on the specific Lorentz factor for this mass. Finally, a nonlinear time rate ${\dot \tau} = e^{(o)}_{\ \mu} dx^\mu/dx^o$ (hereinafter $df/dt \equiv {\dot f}$) of moving material objects in (4) depends on the ratio ${{\dot l}^2 / {\dot \tau}^2} = v^2$. This reverse non-linear relation can be simplified in several subsequent steps through the following equalities to (4): 
\begin {equation}
d\tau  \equiv  dx^o {{1+ U_o m^{-1}_{_N}{\sqrt {1-v^2}} }\over 1-v^iU_i m^{-1}_{_N}{\sqrt {1-v^2} }  } \equiv
  { dt \over {1 - U_oP^{-1}_o - P^{-1}_oU_iv^i}}  \equiv dt{{1 + U_iP^{-1}_o {\dot x}^i}\over {1 - U_oP^{-1}_o}}.
 \end {equation}
 Such time dilatation in (5) by the external four-potential $G^{_N}_\mu = U^{_N}_o/P^{_N}_o$ results in the gravitational Sagnac effect when an observer compares the dynamics of different elementary energy-charges $P_o$ in fields with $U_i\neq 0$. 

Now, one may conclude that the GR time element $d\tau$ in the metric interval (1) and, consequently,  in the physical speed $v=dl/d\tau,$ depends only on four potentials $G_\mu$ for positive probe charges $P_o > 0$. Einstein's theory operates with energy sources of gravitation and describes local interactions between tensor energy-momentum densities of distributed bodies with a global spatial overlap of their GR energies. The potential energy-momentum  $U_\mu \equiv (P_{_N\mu} - m_{_N} {\cal V}_\mu)$ of every probe body contributes to its GR energy content and, therefore, to its  passive energy-charge,  $m_{_N}V_o = P_{o} \equiv E_m$. The ratio $U_\mu/P_{o}$ should be considered in Einstein's gravitation as a universal field four-potential (which is not a covariant four-vector) of active gravitational charges for passive energy-charges. Contrary to Newton's gravitation for masses, Einstein's gravitation is the metric theory for interacting energy systems. The Sun, with the active energy-charge $E_{_M}$, keeps the universal potential $U_\mu / E_m  = $  $\{ -GE_{_M}  r^{-1} ; \ 0 \}$ in the Sun's frame of reference for the passive-inertial energy content $E_m$ of a probe mass $m_{_N}$. Below, we employ the universality of the Sun's potential, $U^{_N}_o/P^{_N}_{o} = - GE_{_M}/r \equiv -r_o/r $, for all planets in our computations for gravitational tests of General Relativity with dilated time (4)-(5) and flat material space filled everywhere by radial gravitational fields $r^{-2}$ and their radial $r^{-4}$ energy-sources.

\section { Flatspace for the planetary perihelion precession}
 

Now we consider the metric tensor (3) for a central gravitational field with a static four-potential, $U_iP^{-1}_o = 0$,  $U_oP^{-1}_o = -  G E_{_M}r^{-1} $, where $E_{_M} =  const \approx Mc^2 $ is the active gravitational energy of the `motionless' Sun (in the moving Solar system).   We use Euclidean geometry for the radial distance $r   \equiv  u^{-1}  $  from the Sun's center of spherical symmetry in agreement with spatial flatness maintained by (3) for any gravitational four-potential $G_\mu$ and its gauge $\partial_\mu\phi$. Let us denote the energy content of a probe mass $m$ in the static central field as a passive energy-charge $P_o = m_{_N}V_o =  m_{_N} {{\sqrt {g_{oo}/(1-v^2)} }} = E_m$. Then, the interval (1) for the passive energy carrier in a central field with $U_i = 0$ takes two equivalent presentations due to (4) and (5),  
\begin {equation}
  ds^2=\left(1-{{G E_{_M}E_m}\over r m}{\sqrt{{{1-{dl^2\over d\tau^2(dl)}}}}}\right )^2 dt^2-dl^2\equiv{{dt^2}\left(1+{{G E_{_M}}\over r}\right)^{-2}}-dl^2,
\end {equation}
where infinite iterations in $d\tau^2(dl)=dt^2\left[1-(GE_{_M}E_m/r m) {\sqrt {{{1-{dl^2/ d\tau^2}}}}} \right ]^2$ over the same $d\tau^2(dl)$ in the Lorentz factor result in ${{dt^2} / [1 + ({{GE_{_M}}/ r})  ]^{2}}$ for the Sun-Mercury potential energy $U_o = - GE_{_M}E_m/r $. Spherical coordinates can be used in (6) for the Euclidean spatial interval  $dl^2 = dr^2 + r^2d\theta^2 + r^2 sin^2\theta d\varphi^2$ in energy-to-energy gravitation.

The non-linear metric solution (6) for the nonlocal energy-source does not coincide with the  Schwarzschild empty-space solution \cite {2} for a central field around a point mass-energy source. Therefore, the Schwarzschild  extension of the SR interval, aside with other empty-space solutions obtained by coordinate transformations, is not the only rotationally invariant solution which GR's tensor formalism can propose for space-time-energy self-organization. Ultrarelativistic velocities,  $v \equiv dl/d\tau \rightarrow 1  $ and ${\sqrt {1-v^2}}\rightarrow 0$, in the Weber-type potential reject  the Schwarzschild singularity in the metric (6). Einstein, `the reluctant father of black holes', very strictly  expressed his final opinion regarding the Schwarzschild radius: `The essential result of this investigation is a clear understanding as to why Schwarzschild singularities do not exist in physical reality' \cite{39}.  
In our view, Schwarzschild's metric solution, and all Birkhoff class solutions for the empty space dogma, originates with {\it ad hoc} modeling of matter in the 1915 Einstein equation in terms of point particles. However, Einstein maintained extended sources for his equation and for physical reality. Below, we prove that the nonlocal matter metric (6) corresponds to the $r^{-4}$ radial energy-charge or source of gravity. Therefore, our analysis denies the empty space paradigm. Non-empty material (energy) space  is in full agreement with Einstein's idea of continuous sources and Newton's `absurd' interpretation of distant attraction through stresses in a material ether (called in 1686 as `God's sensorium').

Our next task is to derive integrals of motion for the Einstein-Grossmann passive material point in a strong central field from the geodesic equations $d^2x^\mu/dp^2=-\Gamma^\mu_{\nu \lambda}dx^\nu dx^\lambda /dp^2$. Nonzero affine connections $\Gamma^\mu_{\nu \lambda}$ for the metric (6) take the following components: $\Gamma^r_{\theta\theta} = -r, \Gamma^r_{\varphi\varphi} = -rsin^2\theta, \Gamma^r_{t t} = dg_{oo}/2dr, \Gamma^\theta_{r\theta} = \Gamma^\theta_{\theta r} = \Gamma^\varphi_{\varphi r} = \Gamma^\phi_{r \varphi }= 1/r, \Gamma^\theta_{\varphi\varphi} = -sin \theta cos \theta,  
\Gamma^\varphi_{\varphi \theta} = \Gamma^\varphi_{\theta \varphi} = ctg \theta, $ and $  \Gamma^t_{t r} =
\Gamma^t_{r t}= dg_{oo}/2g_{oo}dr$, where $g_{oo}$ is the function next to $dt^2$ in the interval (6), $ds^2 = g_{oo}dt^2 - dl^2$.   

By following the verified approach with $\theta = \pi/2 = const$ for the isotropic central field, for example \cite {9},  and by substituting flatspace connections $\Gamma^\mu_{\nu \lambda}$ into GR's geodesic equations, one can define the parametric differential $dp$ and write the following gravitational relations,    
\begin{equation}
{\begin {cases}   
 {g_{oo}dt/dp =  1, dp/ds = g_{oo}dt/ds  = E_m/m = const     \cr \cr
  r^2 {d\varphi/ dp} = J_\varphi = const,  r^2 {d\varphi/ ds} = J_\varphi E_m/m \equiv L = const \cr \cr
(dr/ dp)^2 + (J_\varphi /r)^2 - g^{-1}_{oo} = const ( = - m^2/E^2_m) \cr \cr
   (dr/ds)^2 + (r{d\varphi/ ds})^2 - E_m^2/m^2 g_{oo} = - 1, \cr \cr} 
   \end {cases} }
\end{equation}
  with the first integrals $E_m, m,$ and $J_\varphi$ of the relativistic motion in strong gravitational fields. 

 The last line in (7) is the interval equation $ds^2 = g_{oo}dt^2 - dl^2$ with two integrals of motion $E_m^2/m^2 = g^2_{oo}dt^2/ds^2 $ and $\theta = \pi/2$. Therefore, the scalar invariant (6) is actually the equation of motion  for the constant energy charge $E_m = const $ in a central field with the static Weber-type potential $U^{_W}_o = - (U_o/  r){\sqrt {1-v^2}} \equiv  U_o/(E_m - U_o) = - GE_{_M}/ (r + GE_{_M})$, which is inharmonic for the Laplacian $\nabla^2 U^{_W}_o\neq 0$. Recall that Schwarzschild's curved 3D solution differs from (6) and results in conceptual inconsistencies for the Einstein equation \cite{Nar}.
 We can use (6)-(7) for relativistic motion in strong central fields in order to reinforce the ignored statement of Einstein that Schwarzschild singularities do not exist in physical reality \cite {39}.
There are no grounds for metric singularities either in the interval (6), or in the radial potential $U^{_W}_o(r)$ for $r\rightarrow 0$, because $d\tau / dt = {\sqrt {g_{oo}}} =  r /(r + GE_{_M}) $ is a smooth function. We prove (in the last section) that the non-empty space metric tensor (3), as well as $\nabla^2 U^{_W}_o \neq 0$,  corresponds to the continuous energy-source in the 1915 Einstein equation.

The strong field relations (6)-(7) with the first integrals of motion can be used, for example,  for computations of planetary perihelion precession in the solar system.   
The planet's gravitational energy for the GR energy-to-energy attraction, 
$U_o =  -GE_{_M}E_m r^{-1} \equiv  - r_o E_m u $, where $r_o \equiv GE_{_M}/c^4 = const$ and $u \equiv 1/r $, is small compared to the planet's energy, 
$|U_o| \ll E_m = const$, that corresponds to the non-relativistic motion of a  planet N  (with  $E_m/m = const \approx 1$, $E_m \ll E_{_M}$, and  $v^2 \equiv dl^2/d\tau^2 \ll 1$)
 in the Sun's rest frame, with $U_{i} = 0$.
The GR time element for the planet reads from (6) or (7) as 
\begin {equation}
  ds^2 - dl^2  \equiv d\tau^2(dl)=
dt^2 \left(1 -r_o u {{E_m\over m}} {\sqrt {1 - {{dl^2}\over d\tau^2 (dl)} }}   \right)^2
\approx (1-2r_o u)dt^2 +  r_o u\ dl^2, 
\end{equation}
where we set  $r_ou \ll 1$, $E_m/m = 1$, $dl^2   \ll    d\tau^2 (dl)$,
and $dt^2-d\tau^2 (dl) \ll dt^2$.

The field term with spatial displacement $r_o u dl^2$ on the right hand side of (8) belongs to the time element within the invariant $ds^2$. This displacement corresponds to  the non-linear field nature of time $d\tau (dl) = f (r_o dl/d\tau)$, originating from the Weber-type energy potential $U^{_W}_\mu = U_\mu {\sqrt {1-v^2}}/m$ in (3). Therefore, the invariant (1) cannot be discretionally divided (for fields with such potentials) into separate time and space parts.
There is no departure from Euclidean  space geometry with metrics $dl^2 (\theta = \pi/2)$=$ dr^2 + r^2d\varphi^2 $= $u^{-4}du^2 + u^{-2}d\varphi^2$ in (8), (6), or (1). Again,  a particle's non-linear time with spatial displacement $d\tau (dl)$  differs in (8) from the
proper-time $d\tau_{_O} = {{(1-2r_o u)^{1/2}} }dt$ of the motionless local observer. Displacement  corrections,  $r_o u dl^2/ dt^2$, for the non-relativistic limit are very small compared to the main gravitational corrections, $\mu u$, to Newtonian time rate ${\dot t}^2 \equiv  1  >> 2r_o u >> r_ou dl^2/ dt^2$. However, the dependence of a particle's time element $d\tau^2 $ from spatial displacement $dl^2$ accounts for the reverse value of this time element, $r_o dl^2/d\tau^2$, that is ultimately  a way to restore strict spatial flatness in all covariant relations of Einstein's relativity. Here there is some kind of analogy with electrodynamics where small contributions of Maxwell's displacement currents restore the strict charge conservation in  Ampere's quasi-stationary magnetic law.    

Two integrals of motion from (7) in weak fields, $(1 - 2r_o u) {d t / ds} = E_m/m  $ and
 $r^2{d \varphi / ds}$ = $L$, and (8) result in the equation of a rosette motion for planets,
\begin{equation}
{(1-2r_o u) L^{-2}} + {(1-3r_o u)({u'^2 + u^2})} = {{E^2}L^{-2}m^{-2}},
\end{equation}
where $u'\equiv du/d\varphi$ and $r_o u \ll 1$. Now, (9) may be
differentiated with respect to
the polar angle
$\varphi$,
\begin{equation}
u'' + u - {r_o\over L^2 } = {9\over 2} r_o  u^2 +
3r_o u'' u + {3\over 2}r_o  u'^2,
\end{equation}
by keeping only the largest gravitational terms. This equation may be
 solved in two steps when a non-corrected Newtonian  
 solution, $u_o = r_o L^{-2}(1 + \epsilon cos \varphi)  $,
 is substituted  into the GR correction terms
 at the right hand side of (10).

The most important
correction (which  is sum\-med over century
rotations of the planets) is re\-lated to the `resonance'
(proportional to $ \epsilon cos \varphi$) GR terms.
 We therefore ignore
in (10) all corrections apart from $u^2 \sim 2\mu ^2L^{-4}
\epsilon cos \varphi$
and $u'' u \sim - r_o ^2L^{-4}\epsilon cos \varphi$. Then the  approximate
equation for the rosette motion, $u'' + u - r_o L^{-2} \approx
6r_o^3 L^{-4} \epsilon cos \varphi  $, leads   to the
well known perihelion precession $\Delta \varphi = 6\pi r_o ^2L^{-2}
\equiv 6\pi r_o / a (1-\epsilon ^2) $, which may also be derived through
Schwarzschild's metric approximations with warped three-space, as in 
\cite {6,8,9}. 

It is important to emphasize that the observed result  for a planet 
 in perihelion precession  $\Delta \varphi$ (in the Solar 
  non-empty flatspace with non-linearly warped time by Sun's energy densities) 
 has been derived from the invariant four-interval (1) under  flat
three-space, $\gamma_{ij}=\delta_{ij},$ rather than under `empty but warped' three-space with material peculiarities. 
	
\section {The radar echo delay in flatspace}

The gravitational redshift of light frequency $\omega$ can be considered
 as a direct confirmation that gravity  couples to the energy content
of matter, including the massless photon's energy ${\cal E}_\gamma$, rather than to the scalar mass of the particle.  Indeed, Einstein's direct statement ${\cal E}=mc^2$ for all rest-mass particles is well proved,
but the inverse reading, $m = {\cal E}/c^2$, does not work for electromagnetic waves (with $m = 0$) and requires a new notion, the wave energy-charge ${\cal E}_\gamma/c^2 \equiv m_\gamma \neq 0$.   

 In 1907,  Einstein introduced the Principle of Equivalence for a uniformly accelerated body
 and concluded that its potential energy contributes to the `heavy'
 (passive) gravitational mass \cite {11}. This conclusion of Einstein was 
 generalized in a way that any energy, including light, has `relativistic mass' (the gravitational energy-charge in our terminology). Proponents of this generalization proposed that the photon's energy-charge (`relativistic mass') is attracted by the Sun's mass $M$ in agreement with
the measured redshift  $\Delta \omega / \omega =  \Delta {\cal E}_\gamma / {\cal E}_\gamma
 = \Delta (-m_\gamma GMR_{_S}^{-1}) / m_\gamma c^2 $.
Unfortunately,  the formal application (without the four-vector wave equation) of the `relativistic mass' to zero-mass waves initially resulted in the underestimated light deflection,
  $\varphi = - 2GM/R_{_S}c^2 \equiv - 2r_o/R_{_S}$,
   under the `mechanical free fall' of photons in the Sun's gravitational field \cite {12}.
    In 1917, when Schwarzschild's
option \cite {2} for spatial curvature had been tried for all GR solutions, the new non-Newtonian light deflection, $\varphi = - 4r_o/R_{_S} $, had been predicted due to additional contributions from the supposed spatial curvature in question. Later, all measurements proved this curve-space modification for the gravitational light deflection, providing `experimental evidences' for non-Euclidean three-space in contemporary developments of metric gravitation.  

Below,  we prove that Einstein's GR admits the flatspace concept 
for interpretation of light phenomena in gravitational fields if one properly couples the Sun's rest energy to the photon's wave energy ${\cal E}_\gamma$.
We consider both the radar echo delay  and the gravitational 
 deflection of light by coupling wave's energy-charge to local gravitational potentials. Our purpose is to verify that Euclidean space can perfectly match  
 the known measurements \cite {6,13,14} of light phenomena in the Solar system.
 Let us consider a static gravitational field ($g_i = 0$, for simplicity), where the physical slowness of light, $n^{-1} \equiv  v/c $,
can be derived directly from the covariant Maxwell equations \cite {8}, $n^{-1} 
= {\sqrt  {{\tilde \epsilon} {\tilde \mu}} } =  {\sqrt {g_{oo}} }$. Here and below we associate $g_{oo}$ with the gravitational potential $U_o/P_o$ for a motionless local observer at a given point.  
The measured velocity $v = dl/d\tau_{_O}$,
as well as the observed light  frequency $\omega = \omega_o dt/d\tau_{_O}$,
 is
to be specified  with respect to the  observer's time $d\tau_{_O} = {\sqrt {g_{oo}}
 }dt$. It was Einstein who first correctly associated the light's redshift with different clock rates (of local observers) in the Sun's
gravitational potential \cite {11}.

Compared to the physical speed of light, $dl/d\tau_o = cn^{-1}$, its coordinate speed
\begin {equation}
 {{\dot l}}\equiv{{dl}\over d\tau_{_O}}\times{{d\tau_{_O}}\over dt}={c\over n}\times {\sqrt {g_{oo}}}=c{{g_{oo}}} \equiv c \left(1 +{{r_o}\over r }\right)^{-2} \approx c \left(1-{{2 r_o}\over r}\right) 
\end {equation}
 is double-shifted by the  universal gravitational potential $U_o/P_o$ (or by ${\sqrt {g_{oo}} }^2$)
 in the Sun's gravitational field where $r_o $ =
 1.48 km,  $r_o/ \ll r \approx R_{_S} $.   Notice that both the local physical slowness
  $n^{-1} = {\sqrt {g_{oo}}} $ and the observer time dilation $ d\tau_{_O} / dt
   = {\sqrt {g_{oo}}}$ are  responsible  for
 the double slowness of the coordinate velocity (11), which is relevant to observations of light coordinates or rays under gravitational tests.

A world time delay of Mercury's radar echo reads through  
relation (11) as
\begin {eqnarray}
\Delta t = {2} \int_{l_{_E}}^{l_{_M}} dl \left ( 
{{1}\over {\dot l}} - {1\over c} \right ) 
\approx 
{2\over c} \int_{x_{_E}}^{x_{_M}}  
  {{2r_o dx } \over {\sqrt {x^2 + y^2} }  }   
\approx {4r_o\over c } ln {{ 4r_{_M{_S}} r_{_E{_S}} }\over R^2_{_S} } =
220\mu s,
\end {eqnarray}
where $y \approx R_{_S} = 0.7\times 10^6$ km is the radius of the Sun,
while
$r_{_E{_S}} = 149.5\times 10^6$km and $r_{_M{_S}} = 57.9\times 10^6$km
are the Earth-Sun and Mercury-Sun distances, respectively. It is essential
that we use the Euclidean metric for spatial distance, $r = (x^2 +
y^2)^{1/2}$, between the Sun's center (0,0)
and any point (x, y) on the photonic ray.       
One can measure in the Earth's laboratory only the physical time delay
 $\Delta \tau_{_E}= {\sqrt {g^{_E}_{oo}}} \Delta t$, which practically coincides with the world
  time delay $\Delta t$ in the
Earth's  weak field, {\it i.e.}  $\Delta \tau_{_E} \approx \Delta
t = 220 \mu s$. From here, the known experimental results \cite {6,14} correspond to the radar
echo delay (12), based on strictly flat three-space
within curved space-time with warped time.

\section {Gravitational light bending in non-empty flatspace}

A coordinate angular deflection $\varphi = \varphi_{\infty} - \varphi_{-\infty}$ of a light wave front in
the Sun's gravitational field can be promptly derived in flat space geometry by using the
 coordinate velocity (11) for observations,

\begin {eqnarray}
\varphi\!=\!-2\!\int_{0}^{\infty}\!\!dl{\partial\over
\partial y}{{\dot l}\over c}\!\approx\!-\!2\!\int_{0}^{\infty}\!\!dx
{\partial\over\partial y }{2r_o\over
{\sqrt {x^2 + y^2}}} \nonumber     \\ 
 \approx-4r_o\!\int_0^\infty\!\!{{R_{_S}dx}\over (x^2 + R_{_S}^2)^{3/2}}=-{{4r_o}\over R_{_S}}=-1.75".
\end {eqnarray}
One could also try for massless photons a formal extrapolation of the four-interval equation (1) for a rest-mass particle. However, the most rigorous classical procedure to derive the ray deflection (13) is to apply the verified Fermat principle to light waves. This basic principle of physics should also justify spatial flatness under suitable applications  \cite {Mo}.  

In  agreement with Einstein's original consideration \cite {11}, one may 
relate the vector component $K_{o}$ in the scalar wave equation 
$K_\mu K^\mu =  0$ to the measured  (physical) energy-frequency $\hbar \omega$ of the photon
 ($c K_o = E = \hbar\omega = \hbar\omega_o dt/ d\tau_o, \ \hbar\omega_o$ 
  = $const$). Recall that $P_o$  is also the measured particle's energy in the similar equation, $P_\mu P^\mu = m^2c^4$, for a rest-mass particle.
The scalar wave equation $K_\mu K^\mu = g_{_N}^{\mu\nu}K_\mu K_\nu =  0$
has the following solution for the electromagnetic wave,
 \begin {equation}
\begin {cases}  {K_o \equiv \hbar \omega_o dt/ c d\tau  =
 g_{oo} (K^o - g_i K^i)  \cr \cr
\gamma_{ij}K^iK^j=g_{oo}(K^o - g_i K^i)^2=  
K^2_o/g_{oo}=\hbar^2\omega^2_o dt^2 /c^2g_{oo} d\tau^2 \cr \cr
K^i = \hbar \omega_o dt dx^i / c {\sqrt {g_{oo}}} d\tau dl \cr \cr
 K_i =-[(dx_i / dl) +
  {\sqrt {g_{oo}}}g_i ]\hbar\omega_o dt/ c {\sqrt {g_{oo}}} d\tau, \cr}
\end {cases}
\end {equation} 
with $K_\mu = \{E, - E[(\delta_{ij}dx^j/{\sqrt {g_{oo}} dl} ) + g_i] \}$.

The Fermat-type variations with respect to $\delta \varphi$ and $\delta u$ 
($ r \equiv u^{-1} $, $\varphi$, and  $\vartheta = \pi/2$ 
are the spherical coordinates) for photons in a static  gravitational field are
\begin {eqnarray}
\delta\!\int K_i dx^i = -\delta\!\int\! {{\hbar \omega_o\gamma_{ij}dx^j}
\over
c g_{oo}dl}dx^i 
 =   -{\hbar\omega_o\over c}\delta\!\int  {  { \sqrt {
  du^2 + u^2d\varphi^2   }     (1+r_ou)^2 }\over u^2 } = 0,
\end {eqnarray}
(where $g_{oo}=(1+r_ou)^{-2}$, $g_i=0$,$\gamma_{ij}=\delta_{ij}$, $dl={\sqrt{\delta_{ij} x^ix^j}}$
  = ${\sqrt{ dr^2+r^2d\varphi^2}}$) resulting in  a couple of light ray equations for $r_ou \ll 1$,
\begin {eqnarray}
\begin {cases} 
{(1- 4 r_o u)\left [ \left(  {u'_\varphi} \right )^2 +
u^2    \right] = u_o^2 = const \cr
 u''_{\varphi\varphi} + u =  2r_o u_o^2 \cr} 
 \end {cases}
\end {eqnarray}

Solutions of (16),
 $u \equiv r^{-1}$ $ = u_o sin \varphi $ + $2r_o u^{2}_o 
(1 + cos \varphi)$ and $r_ou_o \approx r_o/R_{S} \ll 1 $,
 may be used for the Sun's weak field. The propagation of  light from
 $r(-\infty) = \infty, \varphi (-\infty)= \pi$
to  $r (+\infty)  \rightarrow  \infty , \varphi (+\infty)
 \rightarrow \varphi_{\infty} $ corresponds to the angular deflection 
 $ \varphi_{\infty} = arcsin [-4r_o R_{_S}^{-1}(1 + cos\varphi_\infty)]$
  $\approx -4r_o/R_{_S} = -1.75''$ from the light's initial direction.
This deflection coincides with (13)
and is in agreement with the known measurements $-1.66''  \pm 0.18'' $,
 for example \cite {6}.

We may conclude that there is no need to warp Euclidean 
three-space for the explanation of the `non-Newtonian' light deflections if one strictly follows Einstein's original approach to light in gravitational fields \cite {11}.
In fact, the massless electromagnetic energy exhibits an inhomogeneous slowness of its physical velocity, 
 $v \equiv
dl/ d\tau_{_O} = c{\sqrt
{g_{oo}}} $, and, therefore, 
a double slowness of the coordinate velocity, $dl/ dt = cg_{oo}$.
This coordinate velocity slowness is related to the coordinate bending of light measured by the observer.  In closing, the variational Fermat's principle supports strict spatial flatness for light in the Solar system.

\section {Geodetic and frame-dragging precessions of orbiting gyroscopes}

Expected precessions of the orbiting gyroscopes in the Gravity Probe B Experiment \cite {Eve} have been calculated by Schiff \cite {Sch} based on the Schwarzschild-type metric for curved and empty 3D space.  We shall revisit the point spin model (criticized below) for the free motion in a central gravitation field. Our spatial interval is always flat due to the intrinsic metric symmetries in the GR four-interval (1) with the metric tensor (3). The tensor formalism can be universally applied to any warped space-time manifold with or without the intrinsic metric symmetries and with or without asymmetrical connections.
 By following Schiff and many other authors we also assume for a moment that the vector geodesic equation,  $dS_\mu / dp =\Gamma^\lambda_{\mu\nu} S_\lambda dx^\nu/dp $, in pseudo-Riemannian four-space with only symmetrical connections, $\Gamma^\lambda_{\mu\nu} = \Gamma^\lambda_{\nu\mu}$, may be applied to the point spin `four-vector' $S_\mu$ with `invariant' bounds $V^\mu S_\mu = 0$ or $S_o = - {\dot x}^iS_i$ for orthonormal four-vectors, 
\begin{equation}
	{dS_i\over dt} = \Gamma^o_{i\nu} S_o {\dot x}^\nu + \Gamma^j_{i\nu} S_j {\dot x}^\nu =
\left (- \Gamma^o_{io} {\dot x}^j  - \Gamma^o_{ik} {\dot x}^k {\dot x}^j + \Gamma^j_{io} + \Gamma^j_{ik} {\dot x}^k \right)S_j. 
\end{equation}
Our flat-space for a strong static field with (3) and  $g^{oi} = 0, g^{oo} = (1 - U_oP_o^{-1})^2 = 1/g_{oo} $, and $g^{ij} = -\delta^{ij} $, would formally maintain an inertial-type  conservation, 
$g^{\mu\nu}S_\mu S_\nu	= ({{S_oS_o}/ g_{oo}}) - \delta^{ij}S_iS_j = ({{\dot x}^i / {\sqrt {g_{oo}}} })( {{\dot x}^j / {\sqrt {g_{oo}}} }) S_iS_j - S^iS_i = ({\vec v} {\vec S})^2 - {\vec S}^2  = const$,  in agreement with Einstein's teaching for a free-falling body. At the same time, Schwarzschild's metric option (curved space) tends to suggest \cite{9,Sch} the non-compensated Newtonian  potential $\phi = - GM/r$ even in the `free fall' equation, $const = g_{Sch}^{\mu\nu}S_\mu S_\nu = ({\vec v} {\vec S})^2 - {\vec S}^2(1 + 2\phi)$.  Therefore, formal applications of the Einstein-Grossmann geodesic relations (derived for spatial translation of material points) to localized spins $S_\mu$ (which are not four-vectors in 4D manifolds with symmetrical affine connections) contradict the GR inertial motion and the Principle of Equivalence.

Our affine connections $\Gamma^\lambda_{\mu\nu} = \Gamma^\lambda_{\nu\mu}$, related to the metric tensor (3), depend only on four field   potentials  $G_\mu \equiv U_\mu P_o^{-1} =  \{U_oP_o^{-1}, U_iP_o^{-1} \} $. This metric tensor has been introduced for the local energy-momentum (2) without any rotational or spin components. Moreover, neither the mechanical part, $K_\mu$, nor the gravitational part, $P_o G_\mu$, in (2) are separately covariant four-vectors in warped space-time with the metric tensor (3).      
Therefore, there are no optimistic grounds to believe that four spin components $S_\mu$ might accidently form a covariant four vector in space-time with  symmetrical connections for translation of the passive   four-vector, $P_\mu \equiv  K_\mu + P_oG_\mu$, for a material point or for energy-momentum densities of distributed bodies.  Nonetheless, we  try by chance these symmetrical  connections for the point spin avenue in question (17) in constant fields  when $\partial_o g_{_{\mu\nu}} = 0$, for simplicity,  
\begin {eqnarray}
\begin {cases} 
{ 2\Gamma^j_{io} = U_jP^{-1}_o\partial_i g_{oo} +
\partial_j(U_i P^{-1}_og_{oo})  - \partial_i(U_j P^{-1}_og_{oo} )
 \cr \cr
 2\Gamma^o_{io}=[(1-U_oP^{-1}_o)^2-U^2_iP^{-2}_o]\partial_i g_{oo} \cr \cr  \ \ \ \ \ \ 
 + P_o^{-1}U_i[\partial_j(U_iP^{-1}_o g_{oo}-\partial_i(U_j P^{-1}_og_{oo} )]\cr \cr
 2\Gamma^j_{ik} = \partial_j (U_iU_kP^{-2}_og_{oo})  - U_kP^{-1}_og_{oo}\partial_i(U_jP^{-1}_o)
- U_iP^{-1}_og_{oo}\partial_k(U_jP^{-1}_o)\cr \cr
2\Gamma^o_{ik} =  [(1-U_oP^{-1}_o)^2 - \delta^{ij}U_iU_jP^{-2}_o ]
[\partial_i (U_kP^{-1}_og_{oo})   + \partial_k (U_iP^{-1}_og_{oo}) ]\cr \cr \ \ \ \ \ \ + U_oP_o^{-1}
[ \partial_i (U_jU_kP^{-2}_og_{oo}) +  \partial_k (U_iU_jP^{-2}_og_{oo}) - \partial_j (U_iU_kP^{-2}_og_{oo})] 
 .\cr} 
 \end {cases}
\end {eqnarray}
One could start with $U_oP_o^{-1} = - GE_{_M}r^{-1} $ and $U_iP_o^{-1} = 2GIr^{-3} [{\vec r} \times {\vec \omega}]_i $ for the homogeneous spherical mass M rotating with low angular velocity, {\it i.e.} ${ \omega} r \ll 1$, $U_iU_i/P^2_o \ll 1$, $E_{_M} \approx M$, and $I = \sum_n m_n {\vec x}_n \times {\vec v}_n \approx  2MR_{_E}^2/5 $ for $R_{_E} <  r$ \cite{8}.  Then, by keeping only linear terms with respect to $U_i/P_o$,  one can rewrite (17) for a slowly rotating gravitational field: 
\begin{eqnarray}
	{dS_i\over dt}  \approx - S_j {\dot x}^j  { \partial_i  ln  {\sqrt {g_{oo}} } } 
	 -  \delta^{jk}S_j {{\partial_i (U_kP_o^{-1}g_{oo})  - \partial_k (U_iP_o^{-1}g_{oo}) } \over 2} \cr 
+ S_jU_jP_{o}\partial_i g_{oo} -  
{\dot x}^j{\dot x^k} S_j{{\partial_i (U_kP_o^{-1}g_{oo}) + \partial_k (U_iP_o^{-1}g_{oo})}\over 2g_{oo}}.
\end{eqnarray}

The last three terms on the right-hand side of (19) are responsible for frame rotation and frame dragging, which vanish for non-rotating centers when ${\vec \omega} \rightarrow 0$ and $U_i/P_o \rightarrow 0 $. 
Precessions of the constant magnitude vector ${\vec J} \approx  {\vec S} - ({{\vec v}}{\vec S}) ({\vec v} + 2 {\vec U}P^{-1}_o )/2$
from the weak-field limit for $g^{\mu\nu}S_\mu S_\nu$ = $[(1-U_oP_o^{-1} )^2 - U_iU_iP_o^{-2}]({\dot x}^jS_j)^2  + 2U_jP_o^{-1}S_j({\dot x}^iS_i) - \delta^{ij}S_iS_j \equiv \delta^{ij}J_iJ_j = const$, if  $(-U_o/P_o)$ $\ll$ 1,  ${\dot x}^i{\dot x}^i \ll 1$, and ${\dot x}^i \approx v^i \approx - \partial_i U_oP^{-1}_o$ in (19),
  \begin{eqnarray}
{dJ_i\over dt} \approx - {J_j\over 2}   
[{v}^j \partial_i (U_o P^{-1}_o)	- {v}^i \partial_j (U_oP^{-1}_o) ]	- {J_j \delta^{jk}\over 2} \left ({{\partial_i U_kP_o^{-1} - \partial_k U_iP_o^{-1}}}  \right)\cr
 + J_j [U_jP_o^{-1}\partial_i (U_oP_o^{-1}) - U_iP_o^{-1}\partial_j(U_oP_o^{-1})]	
,	 
\end{eqnarray}
may be compared with Schiff's non-relativistic prediction  $d{\vec J}/dt$ = $({\vec \Omega}_{geo} +{\vec \Omega}_{fd})\times {\vec J}$ for Gravity Probe B. The second summand at the right hand side of (20), $-J_j  \delta^{jk}({{\partial_i U_kP_o^{-1} - \partial_k U_iP_o^{-1}})/ 2}$$ \equiv $$({\vec \Omega}_{fd} \times {\vec J} )_i$, takes Schiff's answer \cite{Sch} for the frame-dragging precession, 
\begin{equation}
	 {\vec \Omega}_{fd} \approx - {1\over 2}{\vec \nabla} \times  \left( {{2GI\vec r}\over r^3}\times {\vec \omega} \right ) =  {GI\over r^3}\left( { {3{\vec r}({\vec \omega}\cdot {\vec r}) }\over r^2} -  {\vec \omega} \right ).   
\end{equation}

The first and third precession terms in (20) depend on the Earth's radial field $\partial_i (U_oP_o^{-1})$ and they count together geodetic and frame phenomena. These terms provide ${\vec \Omega}_{gf}= - (2^{-1}{\vec v} - {\vec U}P_o^{-1} )\times {\vec \nabla }U_oP_o^{-1}.$  Such a  precession for a point spin model, formally borrowed from the Einstein-Grossmann theory for the passive material point (but without any ill-specified self-rotations), fails to reiterate the already verified de Sitter geodetic precession, ${\vec \Omega}_{geo}= - (3/2) {\vec v}\times {\vec \nabla }U_oP_o^{-1} = 3GM ({\vec r}\times {\vec v})/2r^3 $, of the Earth-Moon gyroscope in the Sun's field and frame, where 
${\vec U} \equiv  \{U_1, U_2, U_2 \} = 0$. Therefore, the point spin should be removed from Einstein's relativity and there is a mathematical argument for this rejection. The point spin approach to GR matter cannot justify that $S_\mu$ is a covariant four-vector in pseudo-Riemannian space-time when  the metric tensor is defined exclusively for invariant translations of a probe material point without self-rotations or for the four-momentum density without a self-rotation notion. Therefore, one cannot place $S_\mu$ into the  Einstein-Grossmann geodesic equation with symmetrical connections. Riemann-Cartan geometries with the affine torsion and asymmetrical connection \cite {Car}  for local spin of all neighboring material points are  still under discussion. 
At the same time, it is well known (Weyl in 1923 and Einstein-Infeld-Hoffmann in 1938, for example \cite{8}) that the inhomogeneous GR time dilation (or different $g_{oo}(r)$ for mass elements rotating over a joint axis) defines a relativistic Lagrangian for the classical non-point gyroscope. Therefore, Einstein's relativity quantitatively explains the de Sitter precession through local non-Newtonian time rates for distributed rotating systems. The non-Newtonian enhanced precession  originates from different GR  time rates in neighboring material points, rather than from a local space curvature in question for the ill-defined GR spin of a point mass.

In our view, the Einstein-Grossmann material point in pseudo-Riemannian space-time can provide a physical basis for densities of distributed matter, but not for self-rotations of points. Point spin models for geodetic and frame-dragging angular drifts of free-falling  gyroscopes cannot be reasonable for GR physics even under formal success  of point-spin approximations for observable data. Any speculations  that the de Sitter geodetic precession of the Earth-Moon gyroscope or that the Mercury perihelion precession have already confirmed non-Euclidean space geometry are against proper applications of GR time dilation by gravitational fields, and against Einstein-Infeld-Hoffmann's physics of slowly rotating systems having finite masses and dimensions. Below, we clarify in more detail why GR sources are always non-point distributions of energy-matter. Einstein's relativity for (nonlocal) energy sources requires non-Schwarzschildian interpretation of gravitational probes, including Lunar-Laser-Ranging and Gravity Probe B data. 

\section {Nonlocal continuous astroparticles and non-empty flatspace in Einstein's and Max\-well's equations }

 Weber's 1846 electric force and 1848 velocity-dependant potential had been inferred from lab experiments with moving non-relativistic bodies \cite {Web}. But why has Weber's electrodynamics not been widely accepted after 1869-71 when it successfully supported the 1847 Helmholtz principle for energy conservation? One of the possible reasons is that Weber's inharmonic potential for Poisson's equation provided practically the same measured forces as the harmonic Coulomb potential for Laplace's equation, while the latter had been `coherently' accepted to reiterate Newtonian fields in Maxwell's electrodynamics. The ultimate price for such a privilege for the Coulombic radial potential is the  empty space paradigm with the harmonic solutions of Laplace's equation  for fields around point charges. Weber's inharmonic radial potential never vanishes completely in the Laplacian resulting in non-empty space distributions of the elementary classical charge.

  At first glance, our Weber-type radial potential $U_o^{_W} = U_o(r){\sqrt {1-v^2}}/m =  U_oP^{-1}_o/ (1 - U_oP^{-1}_o)$ might not work for Newtonian gravitation at all, because this dynamical potential should keep the constant energy charge $E_m \equiv P_o$ under the time-varying Lorentz factor ${\sqrt {1-v^2}}$. However, the velocity-dependant energy vertex $U_o^{_W}(r)$, with $U_oP^{-1}_o = const /r$, rigorously addresses both strong and weak fields for the geodesic motion of passive energy carriers. Below, we prove that Weber-type potential can facilitate  the free fall universality and the Principle of Equivalence for energy charges in strong fields.

   The geodesic 3D force $f_i$ exerted on the energy charge $E_m$ of the  mass $m$  in a constant gravitational field is well defined \cite{8} in General Relativity:        
\begin {equation}
{{f_i}} =  - {E_m} \left \{ \partial_i  \left( - {1\over {\sqrt {g_{oo}}   }}  \right )  +     
v^j[\partial_i (g_{oj}/g_{oo}) -  \partial_j (g_{oi}/g_{oo}) ]  \right\}. 
\end {equation}
We may use the integrals of motion from (7), with ${\sqrt {g_{oo}}}= 1+W_o = 
(1 + GE_{_M}r^{-1})^{-1}$, $ E_m =m{\sqrt {g_{oo}}}/ {\sqrt {1-v^2}} = const$, and $U_iP_o^{-1} = 0$, for strong central fields of the static source of gravitational energy $E_{_M}= const$. Such a source of gravitation results in the radial `geometrical' force, ${\vec f} = E_m {\vec \nabla} (1/{\sqrt {g_{oo}}}) =
E_m{\vec \nabla} (1 + GE_{_M}r^{-1})  $ = 
 $-GE_{_M} E_m{\vec r}/r^3$, which depends on the passive-inertial charge $E_m$   of the probe mass $m$. The well tested  relativistic acceleration ${d{\vec v}/dt} \equiv [{\vec f} - {\vec v} ({\vec v} {\vec f})]/E_m$ also depends on this energy charge (or the energy content $E_m$). Therefore,  the universal gravitational fall is to be expected in strong fields for attraction  between energy charges rather than between masses. In this way, the geodesic equation for probe particles suggests how Einstein's field equation for the source may transfer General Relativity into a self-contained theory without references to Newton's gravitation for masses  or to the empty-space  electromagnetic model for point charges. There is no need  to curve 3D space for the geodesic motion of energy charges if one would like to keep  Einstein's Principle of Equivalence and the energy-momentum tensor source for Einstein's gravitation.

Following Einstein \cite{1}, we assume that the source of a gravitational field is the distributed energy-momentum  density $T_\nu^\mu$ in the 1915 GR equation, $R^\mu_\nu - \delta^\mu_\nu R /2 = 8\pi G T^\mu_\nu$. Hereinafter we discuss only the four `time'-related densities, $T^\mu_o$, with well-defined initial conditions for the differential gravitational equations. It is not an accident that the pseudo-Riemannian  metric (3) with non-empty energy space admits only four independent  Hilbert variables $g^{o\mu}$, while $\delta (g^{ij}) = \delta (\delta^{ij}) \equiv 0$ in flatspace metric formalism. Once we integrate the distributed particle into its spatial field  structure, we have to accept that curved space-time is doubly warped by both the particle-energy density and by its field-energy density. Such a non-dual unification of the continuous particle and its field should lead to the complete geometrization of the bi-fractional ({\it gesamt}, particle + field) non-local carrier of energy with the subsequent vanishing of sources next to the Einstein curvature tensor $g^{\mu\nu}{\hat G}_{o\nu} = {\hat G}^\mu_o \equiv g^{\mu\nu}R_{o\nu} - \delta^\mu_o R /2$ in non-empty flat space, where
\begin {eqnarray}
 \begin {cases} {
 {\hat G}_o^\mu = \delta^\mu_o \Lambda \cr
 {\hat G}_{o;\mu}^\mu \equiv 0.
} \end {cases}
\end {eqnarray}
Here we use the hat sign in order to mark that the Einstein curvature tensor density ${\hat G}_o^\mu$ has incorporated both the continuous particle (or distributed $T^\mu_o$) and its field fractions under the proper introduction of the space-time metric tensor $g_{\mu\nu}$. Peculiarity-free geometrization of non-empty material space of the {\it gesamt} (particle + field) energy carrier with the local  conservation ${\hat G}_{o;\mu}^\mu \equiv 0$ corresponds, in principle, to Newton's ether stresses for gravitating bodies and to Clifford's `space-theory of matter' \cite{Cli}. 

There are no strict requirements to keep the constant $\Lambda$-term for the overlapping distributions of geometrized energy carriers, unless one would like to insert non-metric energies (heat) into the gravitational equation. Therefore, one may put $\Lambda = 0$ in (23) for many of its applications. A world ensemble of overlapping radial energies can be described by the sum of vector contractions $\sum_1^{\infty} u^\nu u_\mu {\hat G}_o^\mu = 0$ of the elementary tensor (23) for one radial carrier. However, natural questions arise: Where does the particle disappear (${\hat G}_o^o = 0$) and what is the energy density of the {\it gesamt} continuous carrier of active and passive gravitational charges, {\it i. e.} integral energies $E_{_M}$?  Zero Einstein tensor curvature in (23) does not mean the absence of local energy-matter. Only a zero Riemann rank-four tensor can justify Minkowski space-time. Otherwise, one may read the local energy balance, $g^{o\nu}R_{o\nu} = R/2 \equiv 8\pi G \epsilon $  in (23) by relating the Ricci scalar curvature $R\equiv g^{\mu\nu}R_{\mu\nu}$, to  joint   particle plus field contributions into the energy density $\mu$ of the {\it gesamt} continuous carrier. 

Let us study non-empty material space filled by the radial particle-field carrier of distributed active, $E^a_{_M}$, 
and passive, $E^a_{_M}$, energies which locally form the symmetrical Ricci tensor $R_{\mu\nu} = R_{\nu\mu} = \partial_\lambda \Gamma^\lambda_{\mu\nu} - \Gamma^\lambda_{\mu\rho}\Gamma^\rho_{\nu\lambda}  + \Gamma^\lambda_{\mu\nu}\partial_\lambda ln {\sqrt {g_{oo}}} - \partial_\mu \partial_\nu ln {\sqrt {g_{oo}}} $ (with $g_{oi} = 0$ and vanishing time derivatives for static states).   The active gravitational energy of a static source warps  $g_{oo}(r) = 1/g^{oo} (r) =  [1 - U_o(r)E_{_m}^{-1}]^{-2}$ for all passive (probe) energies $E_m$ with, as so far, the unreferenced potential energy $U_o(r)$. Only two  warped connections $\Gamma^i_{oo} = \partial_i g_{oo}/2$ and $\Gamma^o_{io} = \partial_i g_{oo}/2 g_{oo}$, when $\partial_o g_{mu} = 0 $ and $U_i = 0$, define $R^o_o = g^{oo}R_{oo} = g^{oo}(\partial_i \Gamma^i_{oo} - \Gamma^j_{oo}\Gamma^o_{oj})$ = $[ -\partial_i^2 ln (g_{oo}^{-1/2})   + (\partial_i ln (g_{oo}^{-1/2}))^2 ] $ and  
$R = g^{oo}R_{oo} + g^{ij}R_{ij} = g^{oo} (\partial_i \Gamma^i_{oo} - \Gamma^j_{oo}\Gamma^o_{oj})$ - $\delta^{ij}(-\partial_j \Gamma^o_{oi}- \Gamma^o_{io}\Gamma^o_{jo}) $ = $2R_o^o$, with 

\begin {equation}
{{R}\over 8\pi G} = {{ g^{oo}R_{oo}}\over 4\pi G} =  {{- \nabla {\vec w} \over 4\pi G}}  + {{\vec w}^2\over 4\pi G} \equiv \epsilon_a + \epsilon_p.
\end{equation}
Here we define the geometrized mass-energy density $(\epsilon_a + \epsilon_p)$ through the scalar Ricci curvature $R$ and introduce the post-Newtonian field intensity ${\vec w } \equiv - \nabla W$ through the inharmonic static potential $W \equiv  - ln (1/{\sqrt {g_{oo}} })$ of distributed particle-field matter.

One may conventianally associate the summand $(- \nabla {\vec w})/ 4\pi G \equiv \epsilon_a$ in (24) with the active (source, yang) energy density  of the {\it gesant} particle-field carrier of active/passive energy-charges,  while the summand  $ {\vec w}^2/ 4\pi G \equiv \epsilon_p$ being associated with the passive-inertial (sink, ying) energy density or vice-versa. These local contributions can be integrated over all material space, defining 
active, $E^a_{_M} = \int \epsilon_a d^3 x$, and passive $E^p_{_M} = \int \epsilon_p d^3 x$ energy-charges of nonlocal GR bodies. 
The Equivalence Principle for the integral charges $E^a_{_M} \equiv E^p_{_M}$ of energy carriers is based on equal densities of their active and passive energies, $\epsilon_a = \epsilon_p$.
Such an  approach to GR charges requests equal particle and field contributions to the Ricci scalar (24).  This specifies the SR referenced metric solutions $ 1/{\sqrt {g_{oo}}} \equiv  1 - U_oE^{-1}_m = 1 + C_1 r^{-1} + C_2$ (3) for the Einstein-Grossmann material point (with the passive energy $E_m = P_o$ without internal rotations) due to the differential equation $[\partial_r ln (1 - U_oE^{-1}_m)]^2 \equiv - r^{-2}\partial_r [r^2\partial_r ln (1 - U_oE^{-1}_m)] $. One constant vanishes ($C_2 = 0$) due to the SR asymptotic $g_{oo} (r\rightarrow \infty) = 1$. In order to find $C_1$ without references to Newton mechanics,              we analyze the active energy distribution of nonlocal sources from (24), 
\begin {eqnarray}
{{E_{_M}}} = {1\over 4\pi G} \int 4\pi r^2 dr \nabla (-{\vec w}) = {1\over G} \int dr \partial_r [ - r^2 \partial_r  ln ({g^{-1/2}_{oo} })  ] \cr 
 = - {r^2\over  G} \partial_r ln (g^{-1/2}_{oo})|^\infty_o = {{r^2 \partial (U_oE_m^{-1})}\over G(1 - U_oE^{-1}_m )} |^\infty_o .  
\end{eqnarray}
  Here we use the metric component $g_{oo}$ with the SR references on the mechanical part $K^{_N}_{\mu}$ of the translation energy-momentum  (2) without internal degrees of freedom. Now, by solving (25) with respect to $U_o(r)$ at $E_m = const$, one can derive the universal attraction law,
      \begin {equation}
       U_o(r) = - {{GE_{_M}E_m}\over r},
       \end {equation}
       for nonlocal carrieres of passive and active energies. Ultimately, we specified for Einstein's SR-GR theory the analytical metric component $g_{oo} = [ 1 + r_o/r ]^{-2} $ and smooth static metric $ds^2 = g_{oo}dt^2 -\delta_{ij}dx^idx^j$, with $r_o \equiv GE_{_M}$. Therefore, the Einstein-Grossmann metric formalism results in the self-contained energy-to-energy gravitation. Again, we did not take Newtonian references for the GR gravitational law (26). Interactions in the self-contained SR-GR theory depend conceptually on attractions  of passive energy charges by active energy charges (and vice-versa), rather than by mutual attractions of constant scalar masses. Any third body can vary the paired interaction energy (26). This approach can quantitatively address Mach's ideas \cite {Mac} regarding  variations of passive-inertial and active gravitational charges ($E_m \neq const$ and $E_{_M} \neq const$             when $\partial_o g_{\mu\nu} \neq 0$) in Newton's 1686 law for gravitation with constant masses. The energy-to-energy attraction (26) may suggest the counting of photons' and neutrino's energy-charges for gravitation on astronomical scales. The static radial field with $g_{oo}^{-1/2} = 1 + r_or^{-1}$ keeps the Gaussian surface flux for the force intensity (22) , $\oint_s ({\vec f}/E_m) d {\vec s} = const $, that corresponds to the energy-charge conservation, $r_o = const$.      
      
            It is essential for the GR geometrization of material flatspace with locally dilated time that the same tensor component $g_{oo}(r)$, which was derived in (6)-(7) for the 1913 Einstein-Grossmann geodesic motion of the passive test body, be followed self-consistently from solution (26) to the gravitational equation (23).   GR energy space is not empty in the Einstein metric curvature  and in physical reality. Contrary to the empty-space model with point sources for Schwarzschild's solution, the Newton-Einsten Universe is filled or charged everywhere with two equal radial energy densities $(- \nabla {\vec w}) / 4\pi G = {\vec w}^2 / 4\pi G  = \epsilon  =    { r^2_o} / 4\pi G r^2(r_o + r)^2$ of static (equilibrium) radial carriers. Our metric geometrization of folded particle-field energies (or {\it yin-yang}, {\it gesamt} fields)  maintains   $r^{-4}$ radial energy elements overlapping on microscopic, macroscopic, and megascopic scales, with equal active (source continuum) and passive (sink continuum) energy densities 
          \begin {eqnarray}
\begin {cases}             
      { \epsilon(r)\equiv {E_{_M} n (r)  } = {  { E_{_M}r_o } /4\pi r^2(r + r_o)^2}  = {\vec w}^2 / 4\pi G =  - \nabla {\vec w}/4\pi G =   \nabla^2 {W}/4\pi G \cr  \cr      
      {\vec w}(r) = - \nabla {W}(r) = -GE_{_M} {\hat {\bf r}}/ r (r + r_o)\cr \cr
       W(r) = - ln [(r + r_o)/r]\cr\cr   
       E_{_M} = \int \epsilon d^3 x = r_o/G      }
       \end {cases}          
     \end {eqnarray}      
           
           The radial particle-field matter (with the post-Newtonian logarithmic  potential W) is in agreement with  the well-known  concerns of Einstein regarding point particles in his 1915 equation: `it resembles a building with one wing built of resplendent marble and the other built of cheap wood'.                         Infinite radial $r^{-4}$ material `tails' of overlapping elementary particles or stars occupy the total Universe in accordance with the `absurd' Newtonian ether and its stresses for gravitation. There are no space regions without matter-energy of continuous radial astroparticles. Why an observed macroscopic body exhibits a finite  volume $V$ with sharp surfaces? Centers of radial symmetry ${\bf R}_i$ of body's $N$ bound atoms with radial scales $r_{oi} \approx  Gm_i \approx 10^{-54} m$ are indeed localized within $V$, which contains the most heaviest portions of elementary non-local energies ${\cal E}_{_V} = \int_{_V} dV (\sum_{i = _1}^{_N} m_in_i({\vec r}-{\vec R}_i) )$. A spatial scale decay of surface atoms is about $10^{-54}m$ that is well below the present limit of space measurements ($10^{-18} m$) and the Planck's length ($10^{-33}m $). At the same time a very small, but finite, portion of body's nonlocal energy $\Delta {\cal E} = (\sum_{i = _1}^{_N} m_i ) - {\cal E}_{_V} <<<  \sum_{i = _1}^{_N} m_i $ is distributed over the entire Universe beyond the perception frames of the body macroscopic volume $V$. In this way, the analytical equilibrium distribution $n ({\vec r}-{\vec R}_i)  = r_o/4\pi ({\vec r}-{\vec R}_i)^2 ( |{\vec r}-{\vec R}_i| + r_o)^2$ of the {\it gesamt} radial field in (27) is a more precise density for matter than the Dirac delta-operator density $\delta ({\vec r}-{\vec R}_i)$.

            Classical Electrodynamics with the delta-operator charge bypassed Newtons's and Clifford's ideas regarding the non-empty space paradigm, as well as Weber's experimental findings of inharmonic radial potentials. However, the exact solution \cite{Bul} to the static Maxwell's equations, 
             \begin {eqnarray}
\begin {cases}             
   { 4\pi \rho (r)  = e r_o / r^2(r + r_o)^2   =  \nabla {\vec D}(r) =  {\vec D}{\vec E} r_e/e \cr\cr
{\vec E}(r) \equiv - \nabla W_e (r) = {e {\hat {\bf r} r_o} /r_e r(r+r_o)} = {\vec D} r_o/r_e  \cr\cr
W_e (r) = (e/r_e) ln [(r+r_o)/r]\cr\cr
 E_e = \int d^3 x \rho W_e = \int d^3 x \rho e/r_e =  \int d^3x {\vec E}{\vec D}/4\pi  =  e^2/r_e , \cr} 
       \end {cases}          
     \end {eqnarray} 
          analytically verifies that the elementary Maxwell electron is distributed over the entire Universe with half of its negative charge, $e  \equiv \int_o^\infty 4 \pi r^2 \rho(r) dr = -e_o$, within the energy radius $r_o \approx r_e = Gm_o = 7\times 10^{-58}m$. Again, the electron's charge density
         $\rho(r) = e n(e)$ is locally proportional to the electron's field energy density, ${\vec E}{\vec D}/4\pi  =  e \nabla {\vec D}/r_e = e\rho(r)/r_e$, like the gravitational particle and field energy densities for the unified non-local carrier. The Maxwell radial charge density $\rho(r)$ is, in fact, the static energy distribution, $\rho(r) e/r_e$, in the constant self-potential $e/r_e$. This particle self-energy identically balances the electric field self-energy and, therefore, electricity does not contribute to the Ricci curvature R of continuous carriers of EM and GR energies. The constant self-potential  
                    $e/r_e$ does not generate self-forces, $\nabla (e/r_e) \equiv 0$, for the equilibrium particle distribution $n(r) = r_o/4\pi r^2 (r + r_o)^2$. Again, the  analytical radial density $\sum_i e_in({\vec r} - {\vec R}_i)$ of nonlocal (astro)electrons around their centers of spherical symmetry at ${\vec R}_i$ can successfully 
replace in Maxwell's equations the Dirac operator density $\sum_i  e_i\delta ({\vec r} -{\vec R}_i)$ for the model interpretation of elementary matter through peculiarities.

Now, we may maintain from (27) that the Earth's radial energy is  distributed over the flat material Universe. Therefore, we expect that the forthcoming Gravity Probe B data will confirm flatspace 
under proper interpretation of spin-orbit and spin-spin frame dragging (with conservation of the spin + orbital angular momentum of non-point GR gyroscopes). This experiment may clarify real affine connections of non-empty material space for the geometrization of rotating matter through the generalized Einstein equation ${\hat G}^\mu_{o} = 0$ for continuous energy carriers. Would this Earth field experiment  be finally explained through  local spatial flatness for translations and rotations, then it may be considered as a basis for non-empty spaces in the classical theory of fields, including electrodynamics of continuous elementary charges and their residual  gravitation with paired (chiral) vector gravitational waves.

\section { Conclusions}

Our main goal is only to reinforce spatial flatness of Einstein-Grossmann's {\it Entwurf} gravitation through energy-driven pseudo-Riemannian geometry, rather than to discuss encountered consequences of the self-contained  SR-GR metric scheme with ${\hat G}^\mu_o = 0$ and ${\hat G}^\mu_{o;\mu} \equiv 0$. To achieve this main goal, we have derived quantitative predictions for Mercury's perihelion precession,  Mercury's radar echo delay,  and the gravitational light deflection by the Sun in strictly flat three-space. The numerical results are well known from the Schwarz\-child empty-space approximation of reality and they were  confirmed in many experiments \cite {6}. Recall that the conventional interpretation of post-Newtonian corrections relies on space curvature near the point gravitational source (including the `point' Sun). On the contrary, our GR analysis allows us to conclude that the non-empty space paradigm  can reinforce the strict spatial flatness in nonlinear GR relations, where $r^{-2}$ Newtonian fields are locally bound with the $r^{-4}$ energy sources. The spatial displacement $dl$ may be referred in Einstein's flatspace relativity as a space interval, while the integral $\int\!dl$ along a space curve does not depend anymore on fields and has a well-defined meaning.    

We attached all field corrections within the GR invariant $ds^2$ to the non-linear time element $d\tau^2$. In other words, gravity may curve specific space-time elements, $d\tau$ and $ds$ for every moving particle, but its space interval $dl$ is always flat and universal. It is not surprising that our approach to relativistic corrections, based on the strong-field equations (7), resulted in Schwarzschild-type estimations, which are also based on similar integrals of motion in the Sun's weak field. However, strong fields in (7) will not lead to further coincidences of numerical solutions with Schwarzschild-type  dynamics in empty space.

Both the Newtonian space interval $dl = {\sqrt {\delta_{i\nu}dx^idx^\nu}} > 0$ and the Newtonian time interval $dt = {\sqrt {\delta_{o\nu}dx^odx^\nu}}$ = $|\pm dx^o| > 0$ 
 are independent from local fields and proper parameters of elementary masses. This absolute universality of world space and time measures is a mandatory requirement for these notions in their applications to different particles and their ensembles. Otherwise, there would be no way to introduce for different observers one  universal ruler to measure  three-intervals and compare dynamics of particles in common world space. For example, it is impossible to measure or compare differently warped four-intervals $ds_{_N}  = {\sqrt {g^{_N}_{\mu\nu}(x)dx^\mu dx^\nu}} $ of different particles. In other words, there is no universal geometry for four-intervals and, therefore, evolution of energy carriers can be observed only in  common three-space which ought to maintain universal  subgeometry  for all material objects.

Satellite tests of the Equivalence Principle for flatspace gravitation may be very useful to verify the equivalence of variable Machian inertial and gravitation energy charges, $P^{in}_o$ = $P^{gr}_o \neq const$, alongside the constancy of scalar masses in time-varying fields. If the Einstein Principle of Equivalence be confirmed by NASA for the energy content of moving bodies, then one should not expect space ripples of flat material space for  energy-to-energy gravitation. 
Our geometrization (23) supports the vector approach to the  gravitational four-potentials $G_\mu$ for GR energy-charges $P_o$ in the global overlap of {\it gesamt} carriers $\sum^{\infty}_i u^\nu u_\mu{\hat G}_o^\mu = 0$.
Paired vector waves form the tensor wave (graviton) according to this gravitational equation for (chiral) matter. Chiral confinement of paired waves in the tensor graviton can explain the period decay of the binary pulsar B1913+16 under the GR energy conservation (without an outward gravitational energy flow). There is the strict conservation of the Gauss energy flux (neglecting EM wave losses) for a two body gravitation system in flat space. Space-time-energy self-organization in Einstein's non-empty space relativity for radial energy charges can be well described without 3D metric  ripples. The internal heat generation from local emission-absorption of chiral GR waves within interacting (inert or live) bodies should be analyzed and tested in practice in parallel with the advanced LISA search of 3D ripples of empty (energy?) space in question.

We expect that measured precessions of the Gravity Probe B gyroscopes will ultimately confirm flatness for non-empty energy space for (chiral) gravitational interactions. Then, the chiral mass origin and the dark energy problem could be analyzed in pure geometrical terms for accelerated time due to a decreasing mass-energy density after the Big Bang fluctuation (toward its complete disappearance at the zero mass-density).
 New tests of paired vector waves in confined tensor gravitons for continuous $r^{-4}$ particles and quantitative justification of (23)-(28) may address many conceptual questions, including the geometrical origin of the chiral inertial energy from two coordinate branches $x^o$ in the real time arrow $dt = |\pm dx^o|$ and the unification of nonlocal inertial and electric complex charges under the Rainich-type conditions\cite {Rai} for gravito-electrodynamics. 
The reinforcement of the universal 3D geometry for proper interpretation of elementary GR and EM radial energies of nonlocal astroparticles in the non-empty  physical space is the very beneficial option for the Einstein-Grossmann metric formalism toward its gauge-invariant integration into the Standard Model.

 
{}

\end {document}